\documentclass[11pt,twoside]{article}

\usepackage{asp2004}
\usepackage{epsf}
\usepackage{epsfig}
\usepackage{psfig}
\usepackage{lscape}

\begin{document}
\title{Spitzer White Dwarf Planet Limits}   %%% Fill in title
\author{F. Mullally, Ted von Hippel, D. E. Winget}   %%% Fill in author names
\affil{Department of Astronomy, 1 University Station, C1400, Austin, TX 78712, USA (fergal@astro.as.utexas.edu)}    
%%% Fill in author affiliations

\begin{abstract} %%% Abstract to run on from here.
We present preliminary limits on the presence of planets around white dwarf stars using the IRAC photometer on the Spitzer space telescope. Planets emit strongly in the mid-infrared which allows their presence to be detected as an excess at these wavelengths. We place limits of  $5~M_J$ for 8 stars assuming ages of $1~Gyr$, and $10~M_J$ for 23 stars.We describe our survey, present our results and comment on approaches to improve our methodology.

\end{abstract}

%%% MAIN BODY OF TEXT GOES HERE. CONSULT "INSTRUCTIONS FOR AUTHORS USING
%%% LATEX2E MARKUP", SECTIONS 2.3-2.6 FOR HELP WITH EQUATIONS, FIGURES,
%%% AND TABLES.

\section{Introduction}   %%% Top level section head (remove "%" symbol)
\begin{figure}[hbt]
    \begin{center}
        \includegraphics[angle=270,scale=0.30]{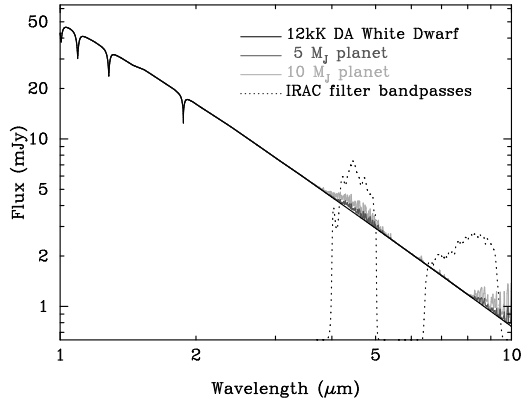}
    \end{center}

    \caption{Model of a 12~kK white dwarf photosphere scaled to a distance of 10~pc with the contribution 5 and 10$M_J$ planet included (dark and light grey respectively). The IRAC photometer sensitivities for Channels 2 and 4 are also shown as dotted lines. \label{irxs} }
\end{figure}

Direct detection of planets around main sequence stars is difficult. Proposed space missions such as Darwin or Terrestrial Planet Finder will require contrast ratios $\sim 10^{-10}$ to detect terrestrial mass planets. As discussed in other contributions to these proceedings, white dwarf stars (WDs) offer a unique advantage in this regard. The luminosity of a 12~kK WD is orders of magnitude less than a progenitor A star. In the mid-infrared the flux from a large ($\sim 5-10 M_J$) planet can rival that of the white dwarf. In Figure~\ref{irxs} we plot a spectrum of a 12~kK DA white dwarf from \citet{Finley97} from 1 to 10$\mu m$ scaled to a distance of 10~pc. We also show the additional flux contribution of an unresolved 5 and $10~M_J$ planetary companion based on models by \citet{Burrows03}. The excess around 4 microns is clearly visible. This excess lies between absorption bands of methane and water. The dashed lines show filter tracings for Channels 2 and 4 (referred to as the 4.5 and 8.0$\mu m$ bands) for the IRAC photometer \citep{Fazio04} on the Spitzer space telescope \citep{Werner04}. By observing an excess at 4.5 compared to other wavelengths we can detect the presence of a planet

\section{Observations}
We obtained infrared photometry of 124 bright ($K < 15.5$) WDs with temperatures from $5-170$~kK at 4.5 and 8$\mu m$. Our observations are discussed fully in Mullally (2006, submitted).
Briefly we observed each star simultaneously in both channels 2 and 4 for a total of 150 seconds. Our flux limit at 8$\mu m$ was somewhat less than 0.1mJy. Each star was processed with version S11.0.4 of the Spitzer pipeline to produce basic calibrated data frames (or bcds). We extracted photometry from each bcd frame using the {\it aper} routine in IDL and compared it to visual photometry from \citet{McCook99} and near infrared photometry from 2MASS.

\section{Models and modeling}
For this work we focused solely on the 77 DA stars with temperatures between 7 and 60~kK (the range of our model grid) in our sample. We created a model spectrum of each WD based on its temperature (and assuming $\log{g} = 8.0$) by linearly interpolating over a grid of models from \citet{Finley97}. To this spectrum we added the flux of a planet model from \citet{Burrows03} for a range of masses from 1 to 25\,$M_J$ and ages of 0.3, 1.0 and 3 Gyr. Performing synthetic photometry of these combined models yielded the model flux at V, 4.5 and 8.0.

\begin{figure}[t]
    \begin{center}
        \includegraphics[angle=270,scale=0.3]{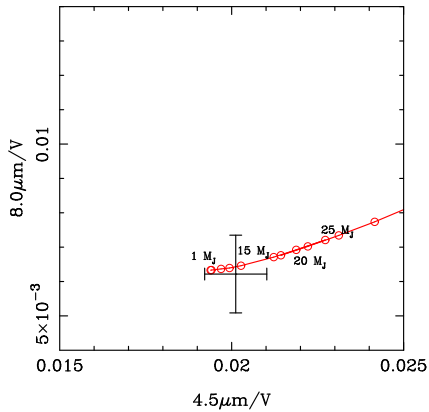}
    \end{center}

    \caption{Flux ratios for a typical star in our survey. The presence of an unresolved low mass companion around the white dwarf increases the observed flux at 4.5 and 8.0$\mu m $ compared to the optical photometry. The 1$\sigma$ limit on planets around this star is better than 15$M_J$.\label{ratio}}

\end{figure}

We then compared the ratio of these fluxes (analogous to colour), to that observed from the DAs in our sample. Figure~\ref{ratio} shows this flux ratio for an example star in our study. As the mass of the planet increases the flux at 4.5, and to a lesser extent 8.0$\mu m$, increases as compared to V. Looking only at 4.5, we determine a mass limit as the mass of planet that would produce an excess equal to the the position of the right hand 1$\sigma$ error bar. For the figure shown above, our limit is $<5~M_J$.

\begin{figure}[hbt]
	\begin{center}
		\includegraphics[angle=270,scale=0.40]{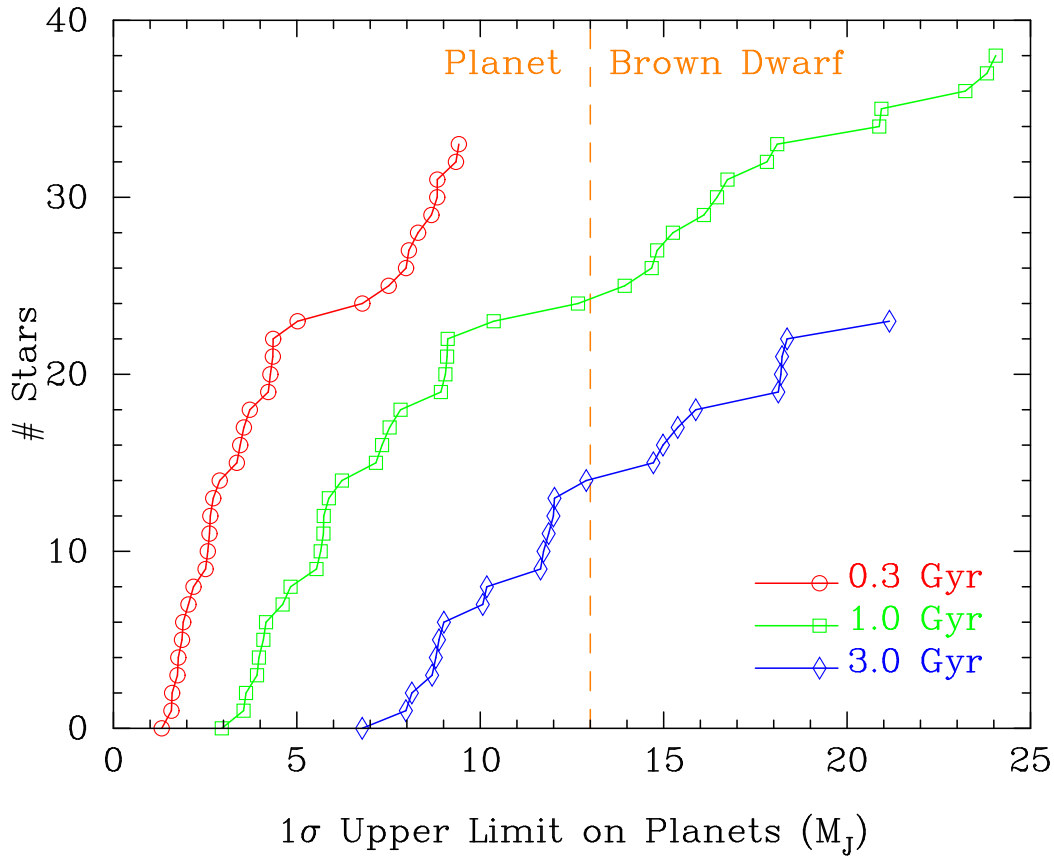}  %Colour version
	\end{center}
    \caption{Cumulative distribution of planet limits from this work for three reasonable planet ages. The circles are for planet models of $0.3$ Gyr, squares are for $1.0$ Gyr and diamonds for $3.0$ Gyr. The vertical dashed line indicates the boundary mass between planets and BDs as defined by the deuterium burning limit. The presence of planets with masses greater than $13 M_J$ and ages less than 1~Gyr has been ruled out around 24 stars in our sample, while planets with masses above 5~$M_J$ have been ruled out for 8 objects. Younger planets are brighter and easier to detect, so have more stringent limits. Planets with masses more than 5~Gyr and ages of 0.3Gyr are ruled out for 24 stars, while the limit is as low as 2 $M_J$ for 8 stars.  
\label{results} }
\end{figure}

Our results are shown in Figure~\ref{results}. The lower axis is the $1\sigma$ upper mass limit, the left axis is the number of systems for which this limit was achieved. Because we do not yet have age information about these systems, we calculated limits for 3 reasonable planet ages. The squares show the number of planets for which a given upper mass limit was achieved for 1~Gyr old planets. There are 8 stars in our survey for which no planets with masses above $5~M_J$ and ages of 1~Gyr were detected, and 23 stars with no planets detected above $10~M_J$. The circles and diamonds show limits for younger (0.3~Gyr) and older (3~Gyr) planets. For a 12~kK WD of an A star progenitor, the age of the system (progenitor lifetime plus WD cooling age) is $\approx$~1.5~Gyr.

For a number of stars, the measured flux ratio lies to blue of the synthetic photometry of an unaccompanied WD. This corresponds to an observed flux deficit at 4.5 relative to V. This causes the $1\sigma$ limit on planetary companions to be unreasonably low, and these stars were excluded from our result. We placed limits on a total of 39 stars.

\section{Improvements}
As discussed elsewhere in this volume, comparing Spitzer fluxes to V band measurements can lead to false deficit measurements, which may explain the anomalously low limits we encountered. A better approach would be to measure the excess over a model normalized to all available visible and near-IR photometry, or to directly fit the observed spectral energy distribution to a grid of white dwarf + planet models, which we intend to do in the near future. Another concern is the recent paper of \citet{Patten06} who found that T dwarves show a smaller than predicted flux in IRAC's Channel 2. They attribute this to larger than expected CO absorption, and may reduce the sensitivity of our survey.

\section{Conclusion}
With their low luminosity relative to their progenitors, WDs over a unique opportunity to directly detect the flux from a companion planet. In this first-look survey with the Spitzer Space Telescope, we surveyed a large number (124) white dwarf stars, but found no convincing evidence of a planet. In this work we make a first cut at placing limits on planets using this data. We find that for planets between the ages of 0.3 and 1~Gyr we can rule out the presence of planets with masses larger than between 8 and 16~$M_J$ for 30 systems.

\acknowledgements %%% Text of acknowledgements runs on after this command.
This work is based on observations made with the Spitzer
Space Telescope, which is operated by the Jet Propulsion Laboratory,
California Institute of Technology under NASA contract 1407. Support for
this work was provided by NASA through award project NBR 1269551 issued
by JPL/Caltech to the University of Texas. This work is performed in part under contract with the Jet Propulsion Laboratory (JPL) funded by NASA through the Michelson Fellowship Program. JPL is managed for NASA by the California Institute of Technology. One of us (FM) wishes to thank the Royal Astronomical Society for a generous grant. Support of the Royal Astronomical Society as well of the American Astronomical Society and the National Science Foundation in the form of an International Travel Grant, which enabled one of us (FM) to attend this conference.

%%% THE BIBLIOGRAPHY
%%%
%%% CONSULT SECTION 3 OF "INSTRUCTIONS FOR AUTHORS" FOR HOW TO USE NATBIB.
%%% AUTHORS ARE ENCOURAGED TO USE EITHER THE "THEBIBLIOGRAPY" ENVIRONMENT
%%% BY UNCOMMENTING (DELETING THE "%" SYMBOL) THE COMMANDS BELOW, OR BY
%%% USING THE BIBTEX ENVIRONMENT. TO FIND OUT WHICH IS APPLICABLE TO YOUR
%%% CONTRIBUTION, CONSULT THE VOLUME EDITORS FOR YOUR PROCEEDINGS.
%%%

%\small
%\bibliographystyle{apj}
%\bibliography{sirtf,planet,wdwarf}
%\normalsize
%%\setlength{\itemsep}{0ex} % Put this line in the bbl file

\end{document}